\PassOptionsToPackage{unicode}{hyperref}
\PassOptionsToPackage{hyphens}{url}
\documentclass[
]{article}

\usepackage{etoolbox}
\AtBeginEnvironment{abstract}{\LARGE}

\usepackage{amsmath,amssymb}
\usepackage{lmodern}
\usepackage{iftex}
\ifPDFTeX
  \usepackage[T1]{fontenc}
  \usepackage[utf8]{inputenc}
  \usepackage{textcomp} 
\else 
  \usepackage{unicode-math}
  \defaultfontfeatures{Scale=MatchLowercase}
  \defaultfontfeatures[\rmfamily]{Ligatures=TeX,Scale=1}
\fi
\IfFileExists{upquote.sty}{\usepackage{upquote}}{}
\IfFileExists{microtype.sty}{
  \usepackage[]{microtype}
  \UseMicrotypeSet[protrusion]{basicmath} 
}{}
\makeatletter
\@ifundefined{KOMAClassName}{
  \IfFileExists{parskip.sty}{%
    \usepackage{parskip}
  }{
    \setlength{\parindent}{0pt}
    \setlength{\parskip}{6pt plus 2pt minus 1pt}}
}{
  \KOMAoptions{parskip=half}}
\makeatother
\usepackage{titling}
\usepackage{etoolbox}
\usepackage{tgtermes}
\usepackage[T1]{fontenc}
\usepackage{amsmath}
\usepackage{newtxtext} 
\usepackage{helvet}
\usepackage{lmodern}
\usepackage{kpfonts} 
\usepackage{textcomp}
\usepackage{sectsty}
\usepackage{xcolor}
\usepackage{longtable,booktabs,array}
\usepackage{calc} 
\usepackage{etoolbox}
\makeatletter
\patchcmd\longtable{\par}{\if@noskipsec\mbox{}\fi\par}{}{}
\makeatother
\IfFileExists{footnotehyper.sty}{\usepackage{footnotehyper}}{\usepackage{footnote}}
\makesavenoteenv{longtable}
\usepackage{graphicx}
\makeatletter
\def\maxwidth{\ifdim\Gin@nat@width>\linewidth\linewidth\else\Gin@nat@width\fi}
\def\maxheight{\ifdim\Gin@nat@height>\textheight\textheight\else\Gin@nat@height\fi}
\makeatother
\setkeys{Gin}{width=\maxwidth,height=\maxheight,keepaspectratio}
\makeatletter
\def\fps@figure{htbp}
\makeatother
\ifLuaTeX
  \usepackage{selnolig}  
\fi
\IfFileExists{bookmark.sty}{\usepackage{bookmark}}{\usepackage{hyperref}}
\IfFileExists{xurl.sty}{\usepackage{xurl}}{} 
\hypersetup{
  hidelinks,
  pdfcreator={LaTeX via pandoc}}

\author{}
\date{}

\makeatletter
\renewcommand{\maketitle}{
  \begin{center}
    \hrule height 1.5pt
    \vspace{0.99em}
    {\LARGE\bfseries\fontfamily{ptm}\selectfont \@title}
    \vspace{0.99em}
    \hrule height 1.5pt
  \end{center}
}
\makeatother

\title{Highly accurate and efficient deep learning paradigm for
full-atom protein loop modeling with KarmaLoop}

\begin{document}

\maketitle


\vspace{0.3cm}
\begin{center}
\textbf{Tianyue Wang\textsuperscript{1\#}, Xujun Zhang\textsuperscript{1\#},
Odin Zhang\textsuperscript{1\#}, Peichen Pan\textsuperscript{1,*}, Guangyong Chen\textsuperscript{2,*},
Yu Kang\textsuperscript{1,*}, Chang-Yu
Hsieh\textsuperscript{1,*}, Tingjun Hou\textsuperscript{1,*}}

\vspace{0.3cm}
\textsuperscript{1}Innovation Institute for Artificial Intelligence in
Medicine of Zhejiang University, College of Pharmaceutical Sciences,
Zhejiang University, Hangzhou 310058, Zhejiang, China

\textsuperscript{2}Zhejiang Laboratory, Hangzhou 311100, Zhejiang, China
\end{center}

\renewenvironment{abstract}
 {\begin{center}
  \Large\textbf{Abstract}\\[1.5ex]
  \normalfont\normalsize
  \begin{minipage}{0.85\textwidth}}
 {\end{minipage}\end{center}}

\begin{abstract}

\noindent \normalsize Protein loop modeling is the most challenging yet highly non-trivial
task in protein structure prediction. Despite recent progress, existing
methods including knowledge-based, \emph{ab initio}, hybrid and deep
learning (DL) methods fall significantly short of either atomic accuracy
or computational efficiency. Moreover, an overarching focus on backbone
atoms has resulted in a dearth of attention given to side-chain
conformation, a critical aspect in a host of downstream applications
including ligand docking, molecular dynamics simulation and drug design.
To overcome these limitations, we present KarmaLoop, a novel paradigm
that distinguishes itself as the first DL method centered on full-atom
(encompassing both backbone and side-chain heavy atoms) protein loop
modeling. Our results demonstrate that KarmaLoop considerably
outperforms conventional and DL-based methods of loop modeling in terms
of both accuracy and efficiency, with the average RMSD improved by over
two-fold compared to the second-best baseline method across different
tasks, and manifests at least two orders of magnitude speedup in
general. Consequently, our comprehensive evaluations indicate that
KarmaLoop provides a state-of-the-art DL solution for protein loop
modeling, with the potential to hasten the advancement of protein
engineering, antibody-antigen recognition, and drug design.\\

\end{abstract}
\textbf{\Large 1  Introduction}

Loops are irregular segments in protein structures that often link two
regular secondary structures such as alpha helices or beta
sheets\textsuperscript{1,2}. These structures are typically situated
near the surface of proteins\textsuperscript{3}, positioned in a way to
allow them to facilitate numerous essential biological
functions\textsuperscript{4}, including regulation of enzyme
activity\textsuperscript{5-8}, protein-protein interaction
(PPI)\textsuperscript{9} and protein-ligand
recognition\textsuperscript{10-12}. Given their high flexibility, loops
heavily influence the overall protein dynamics\textsuperscript{13},
rendering loop prediction one of the most challenging yet indispensable
tasks in protein modeling\textsuperscript{14}. Furthermore, since over
half of the experimentally determined structures in the RCSB Protein
Data Bank (PDB)\textsuperscript{15} contain missing
loops\textsuperscript{4}, accurate modeling of loop structures is of
high significance. Particularly, one of the specific tasks in protein
loop modeling is the prediction of antibody complementarity determining
region (CDR) H3 loops. Compared with other CDRs, H3 has much higher
flexibility, rendering it the most challenging task in antibody
modeling\textsuperscript{16}. To address this issue, various
computational methods including traditional methods (knowledge-based,
\emph{ab initio} and hybrid) and deep learning (DL) methods have been
proposed.

Knowledge-based methods, such as FREAD\textsuperscript{17},
LoopIng\textsuperscript{18}, SuperLooper\textsuperscript{19}, and
DaReUS-Loop\textsuperscript{20}, rely on template repositories collected
from pre-existing protein structures and perform modeling based on
certain standards, such as structural similarity. For instance, FREAD
performs database searching using the four rule-based filters based on
anchor C\textsubscript{$\alpha$} separations and the environmentally
constrained substitution score. LoopIng selects templates from the
database by taking advantage of both sequence- and geometry-related
features. SuperLooper provides the first online server for modeling
globular and membrane protein loops, where the loop candidates are
chosen by the standard of the RMSD of the stem atoms. DaReUS-Loop uses
fragments from remote or unrelated proteins to complete loop modeling
and filters by sequence similarity and conformational profile. These
methods are widely used in loop modeling, particularly in cases where a
comparable fragment is available in the template database. However, the
modeling performance of these methods for novel structures (especially
for long loops) would drop sharply without proper references, suggesting
the accuracy of these methods is severely limited by the available
templates\textsuperscript{21,22}.

\emph{Ab initio} methods, including DISGRO\textsuperscript{1},
LEAP\textsuperscript{23}, Cyclic coordinate descent
(CCD)\textsuperscript{24,25}, robotics-based kinematic closure
(KIC)\textsuperscript{26} and next-generation KIC
(NGK)\textsuperscript{27}, offer advantages over knowledge-based methods
due to their independence from templates and the ability to explore a
broader conformational space. However, they need more computational
resources than knowledge-based methods, and their computational cost
increases exponentially with the loop length\textsuperscript{21}.
DISGRO, for instance, employs a chain growth sequential Monte Carlo
sampling strategy for loop modeling, which has been described as an
efficient computational method and particularly effective for modeling
loops with 10-17 residues. CCD minimizes the distances between the
backbone atoms of the C-terminal moving anchor and the corresponding
atoms in the fixed anchor by adjusting one dihedral angle at a time. KIC
utilizes a robotics-inspired local loop reconstruction method for loop
modeling, achieving sub-angstrom accuracy. Among \emph{ab initio}
methods, NGK has been commonly regarded to be the
state-of-the-art\textsuperscript{20,21}, as it combines intensification
and annealing strategies and yields a four-fold improvement over
standard KIC.

Hybrid methods, such as CODA\textsuperscript{28} and
Sphinx\textsuperscript{22}, integrate both \emph{ab initio} and
knowledge-based methods to improve performance\textsuperscript{20}. CODA
utilizes a combination of the knowledge-based approach FREAD and the
\emph{ab initio} method PETRA\textsuperscript{29} for generating a
consensus prediction that should satisfy a set of rule-based filters.
Sphinx combines FREAD and an in-house \emph{ab initio} method that first
selects potential fragments shorter than the length of loop to be
modeled. Then, the \emph{ab initio} method adjusts the length of the
fragments to generate accurate decoys of the desired length.

With the continual development of computing
resource\textsuperscript{30}, deep learning (DL) has become increasingly
popular and has demonstrated its potential in predicting protein
structures. For instance, AlphaFold\textsuperscript{31} that
incorporates biological and physical information and utilizes
multi-sequence alignments (MSA) to design a DL algorithm, achieves
atomic-level accuracy and ranks first in the 14th Critical Assessment of
Structure Prediction (CASP) competition. Likewise,
RoseTTAFold\textsuperscript{32} utilizes a three-track neural network
that simultaneously considers protein sequences, amino acid
interactions, and structures to attain accuracy close to
AlphaFold\textsuperscript{33}. However, both methods often face
significant challenges in accurately modeling loop
regions\textsuperscript{33,34}. The first DL method for loop modeling
was proposed in 2019, which employs Convolutional Neural Networks
(CNN)\textsuperscript{35-39} on the distance matrix to predict the
pairwise distances of loop atoms for conformation reconstruction.
However, this method exhibits lower predictive accuracy compared to NGK
on multiple test datasets.

More recently, DeepH3\textsuperscript{40} was introduced as an antibody
CDR H3 loop modeling method\textsuperscript{41} using CNN to predict the
inter-residue distances and orientations from antibody heavy and light
chain sequence. Furthermore, DeepAb\textsuperscript{42}, the updated
version of DeepH3, has demonstrated improvements with the inclusion of a
pre-trained antibody sequence model and an interpretable attention
mechanism that focuses on important residue pairs to accurately predict
the structures of antibodies. ABlooper\textsuperscript{43} is another
approach that utilizes the E(n) Equivariant Graph Neural Networks
(EGNN)\textsuperscript{44} architecture to predict the backbone atoms of
antibody loops, but it is slightly less accurate than DeepAb.

So far, both traditional and DL-based methods have been unable to
achieve high accuracy and efficiency simultaneously. Moreover, most
proposed methods have focused solely on the heavy atoms of the backbone
within the loop region\textsuperscript{1,24,45-49}. Nonetheless, it is
crucial to predict the conformation of side chains accurately,
especially in the context of protein-ligand
docking\textsuperscript{50-52}, which highlights the significance of
executing meticulous full-atom modeling of loops.

For the reasons above, we present a novel end-to-end DL paradigm named
KarmaLoop, which is designed for accurate and efficient full-atom
prediction of protein loops. Figure 1A depicts an example of a loop
modeled by KarmaLoop which starts with a random initialization of the
loop coordinates and then predicts an accurate loop structure. KarmaLoop
(Figure 1B) deploys GNN architectures comprising two encoders (i.e.,
Graph Transformer\textsuperscript{53-55} (GT) and Geometric Vector
Perceptrons\textsuperscript{56} (GVP)) for intramolecular interaction
encoding, a Mixture Density Network (MDN) block for obtaining confidence
score, and an EGNN block for generating the conformation. The
innovations are as follows: (1) protein residues and atoms are embedded
hierarchically, and atom features are aggregated into the corresponding
residues to attain a fusion representation (Figure 1C); (2) distribution
of the minimum distance between loop and non-loop nodes fitted by the
MDN block introduces an inductive bias to the shared encodes which can 
facilitate pose generation and confidence prediction; (3) self-attention-based EGNN 
with fully-connected interaction graphs is deployed to enable fast coordinates prediction;
\includegraphics[width=4.8in,height=4.4in]{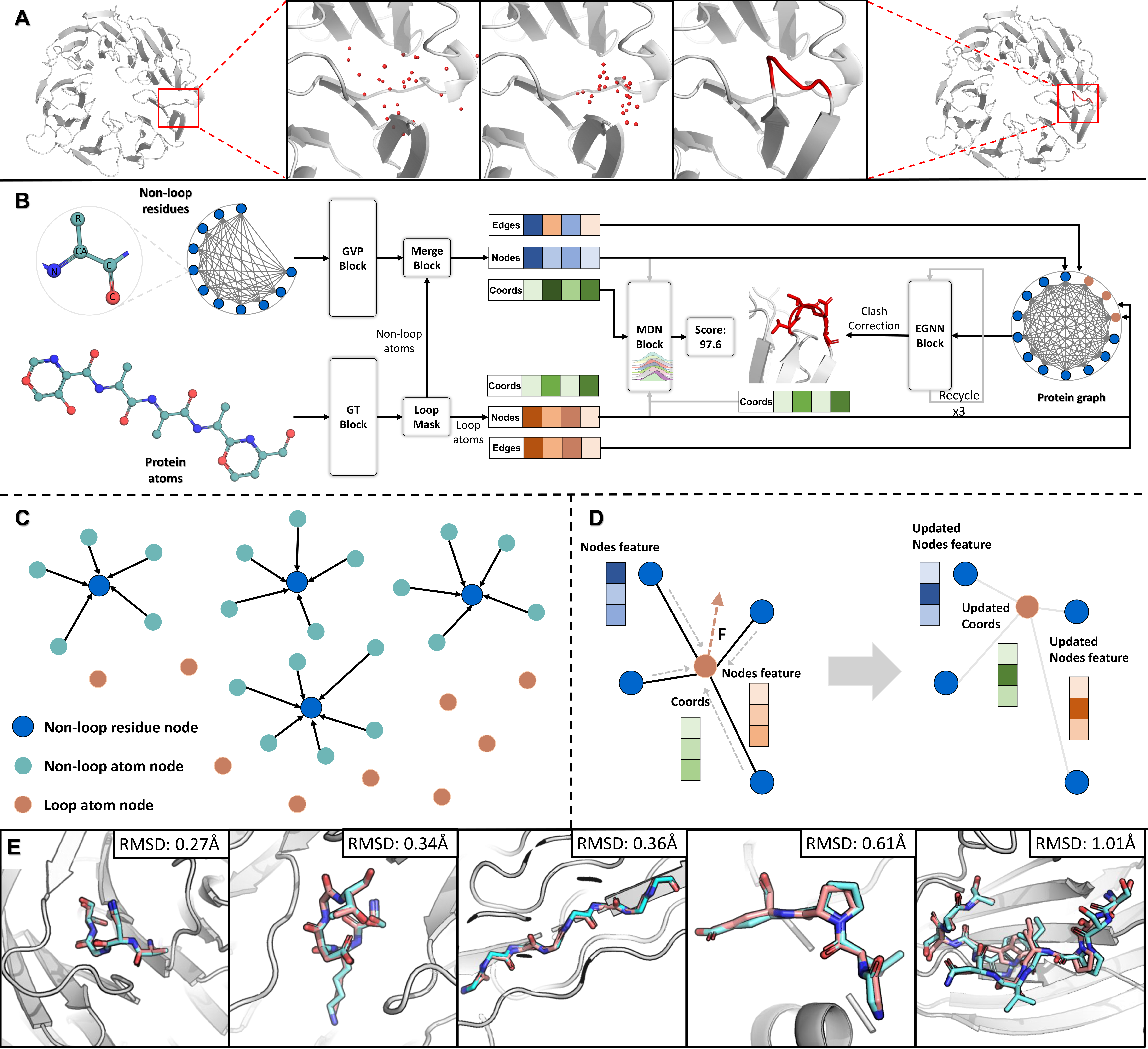}
{\footnotesize Figure 1.The workflow of KarmaLoop for loop modeling.\\}

 (4) a clash correction module is implemented to assist in forming 
chemically plausible conformations.

KarmaLoop has been evaluated on multiple 
benchmark datasets, includingthe general protein loop benchmark CASP13+14 
(the combination of CASP13\textsuperscript{57} and CASP14\textsuperscript{58}), the CASP15
dataset and the antibody H3 loop benchmark (RosettaAntibody and
therapeutic benchmark used in DeepAb). In terms of the general protein
loop modeling capability, KarmaLoop outperforms all the tested
traditional and DL-based methods on both accuracy and efficiency, 
with the medium full-atom root-mean-square deviations (RMSDs) of 1.02 Å and
1.03 Å, the average RMSDs of 1.36 Å and 1.49 Å, the success rates of
81.9\% and 83.9\%, and the speeds of 0.047 seconds (s) and 0.049s per
task on the CASP13+14 and CASP15 benchmark datasets, respectively.
Several successful predictions of antibody CDR H3 loops are shown in
Figure 1E, where KarmaLoop achieves the average full-atom and medium
RMSDs of 2.09 Å and 2.24 Å respectively, compared with the baseline
methods with the best performance (average and medium RMSDs of 3.40 Å
and 3.74 Å, respectively). Additionally, when using the DeepAb-modeled
antibody structures as the input, KarmaLoop could further refine the
conformations of H3 loops and increase the accuracy by 13.14\% and
17.35\% on the medium and average RMSDs, respectively. Hence, we
proposed a universal protocol for precise antibody modeling by employing
DeepAb to first predict an overall antibody structure, followed by
KarmaLoop to refine the H3 loops. In summary, KarmaLoop has shown a
remarkable superiority (over prior arts) in predicting both general
loops and antibody H3 loops across most loop lengths.\\

\textbf{\Large 2  Results and Discussion}

\textbf{\large 2.1  Prediction accuracy for general protein loops}

The accuracy of loop modeling is of utmost importance for many practical
applications. To validate the capability of our model, 6 widely used
methods including two DL-based methods (AlphaFold and RoseTTAFold), one
knowledge-based method (FREAD), and three \emph{ab initio} methods
(DISGRO, NGK, and Rosetta model\_missing\_loop (RML)) were employed for
comparison with KarmaLoop in a comprehensive benchmark. The full-atom
RMSDs were calculated for each method with the AlphaFold and RoseTTAFold
predicted structures being aligned before computation.

To perform quantitative comparisons with the other methods, empirical
Cumulative Distribution Function (eCDF) was selected to describe the
proportion of samples at given RMSD values. We defined a task as
successful if the RMSD of a given sample falls below a specified
threshold. Figure 2A\&B demonstrates that when the RMSD threshold was
set to 2 Å, KarmaLoop exhibited the highest success rates of 83.91\% and
81.80\%, respectively, on the CASP13+14 and CASP15 benchmark datasets,
yielding significant improvements by 28.09\% and 19.90\% over the next
best method (i.e., AlphaFold, 55.82\% and 61.90\%). Even adopting a
stricter RMSD standard of 1 Å, KarmaLoop (46.71\% and 49.38\%) still
outperformed the other methods, ranging from RoseTTAFold (2.55\% and
1.26\%) to NGK (21.69\% and 23.26\%). Of note, KarmaLoop achieved the
success rates of 97.48\% and 94.76\%, respectively, on the two benchmark
datasets when the threshold was set to 4 Å.

As some baseline methods may generate multiple conformations for a given
loop sample, KarmaLoop was further evaluated in the context of
generating multiple conformations (KarmaLoop generated one conformation
while AlphaFold and RoseTTAFold produced 5 conformations and the other
baseline methods outputted 10 conformations). It is important to note
that, regardless of the number of conformations generated for a given
task, a successful prediction is acknowledged when the lowest RMSD is
less than 2 Å. As shown in Figure 2C, KarmaLoop achieved the best
performance and improved the success rates by 13.27\% and 9.94\%,
respectively, on the two benchmark datasets, compared with the best
model NGK with the success rates of 70.64\% and 71.36\%, respectively.
The performance of RoseTTAFold was the worst with the success rates of
only 39.38\% and 30.58\%, respectively.

To directly compare the performance, the distribution of each method was
plotted in Figure 2D. KarmaLoop showed the lowest medium RMSDs on the CASP13+14
and CASP15 datasets,with 1.03 Å and 1.02 Å, respectively, as compared to the other methods,
and 
\includegraphics[width=4.8in,height=5.29in]{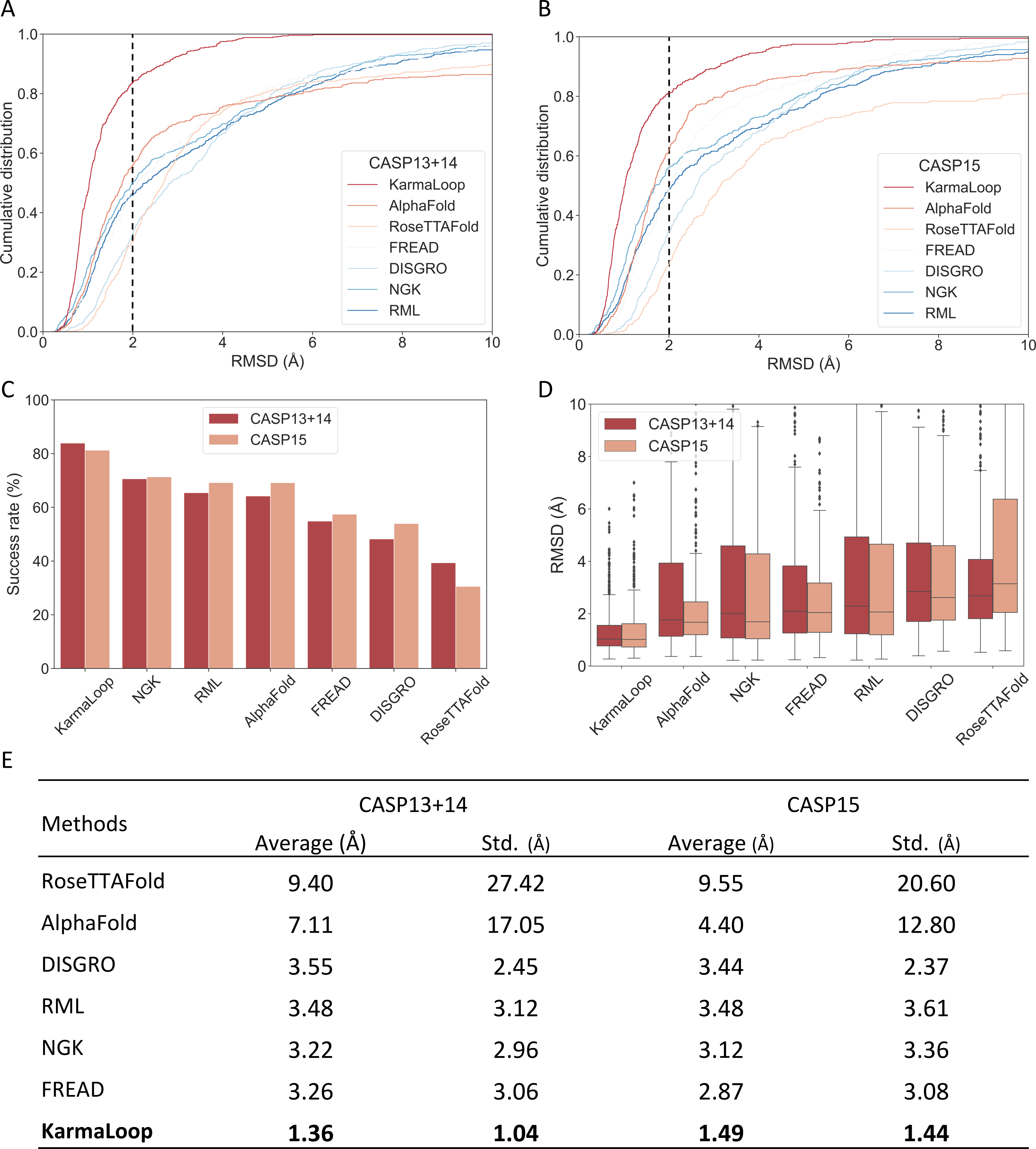}

{\footnotesize Figure 2. Performance of KarmaLoop and protein loop modeling
methods on the CASP dataset. (A) Empirical cumulative distribution of
RMSD of the tested methods on (A) the CASP13+14 dataset and (B) the
CASP15 dataset, and the dashed line indicates the 2Å RMSD cut-off. (C)
The success rate was calculated in the 2Å RMSD threshold, with KarmaLoop
generating only \textbf{one} conformation, AlphaFold and RoseTTAfold
generating \textbf{5} conformations and the other methods generating
\textbf{10} conformations. (D) The RMSD distribution of the tested
methods, and the red boxes indicate the CASP13+14 dataset and the salmon
boxes indicate the CASP15 dataset. (E) The average and standard
deviation (Std.) of the RMSDs of all the tested methods.\\}

showed much lower RMSDs than the
second-best method AlphaFold (1.77 Å and 1.67 Å). Besides, KarmaLoop
achieved the average RMSD of 1.36 Å (±1.04 Å) and 1.49 Å (±1.44 Å),
holding a significant leading margin of 57.76\% and 52.24\% over the
next best baseline method NGK on the CASP13+14 and CASP15 datasets
(Figure 2E).

\textbf{\large 2.2  The impact of loop length on accuracy}

\includegraphics[width=4.8in,height=3.03in]{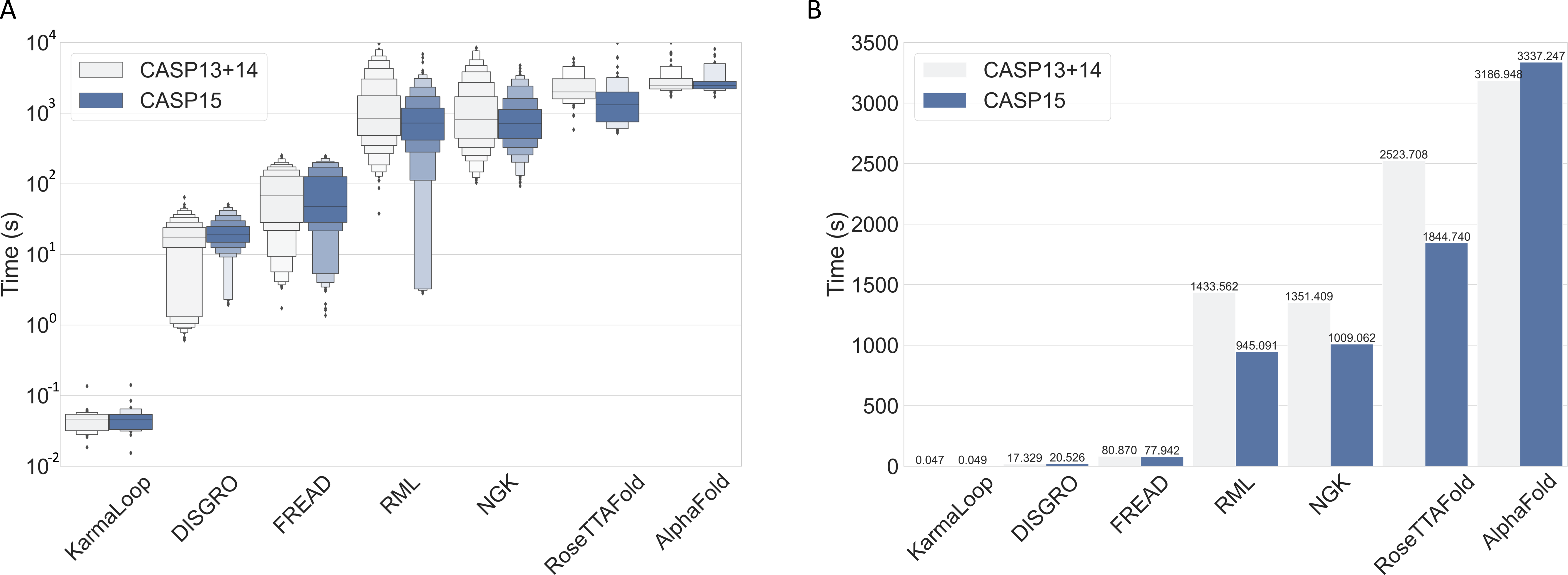}

{\footnotesize Figure 3. Performance of KarmaLoop and the other tested methods
across various loop lengths. The performance tested on (A) the CASP13+14
benchmark dataset and (B) on the CASP15 dataset.\\}

Long loop prediction has been an enduring challenge in loop
modeling\textsuperscript{21}. Therefore, we investigated the influence
of loop length on prediction accuracy. As shown in Figure 1E, KarmaLoop
could generate satisfactory conformations of loops with varying lengths.
Figure 3A\&B displayed the RMSDs of each loop length in the CASP13+14
and CASP15 datasets. Our results indicated that KarmaLoop outperformed
the baseline methods for the majority of length cases (15/23 and 18/25
on the CASP13+14 and CASP15 datasets), and maintained a relatively
consistent performance across different loop lengths. The RMSDs for most
loops shorter than 10 residues were below 2.0 Å. As the loop length
increased to 20 residues, the majority of the samples exhibited RMSDs
less than or equal to 4.0 Å. Notably, the RMSDs were below 10 Å for
almost all loop lengths, even for extremely long loops. The observed
increase in RMSD with loop length may be attributed to the fact that the
training dataset predominantly consisted of short loops (residues$\leq$10),
and long loops (particularly those over 20 residues) were only
represented by 0.058\% of all the samples.

\textbf{\large 2.3  Evaluation of loop modeling efficiency.}

Computational efficiency is also a key factor in loop modeling. In this
section, the time consumption of each method was evaluated. KarmaLoop
was tested on a Tesla V100S GPU, traditional methods (FREAD, DISGRO, NGK
and RML) were run on a single core Intel(R) Xeon(R) Gold 6240R CPU @
2.40GHz, and AlphaFold and RoseTTAFold were evaluated in parallel with
20 cores on the Intel(R) Xeon(R) Gold 6240R CPU @ 2.40GHz and a Tesla
V100S GPU. Figure 4A shows the distribution of time spent for each
method, illustrating the remarkable efficiency of KarmaLoop in terms of
time cost on each task ranging from 0.015s to 0.14s. Contrastively, the
time cost of DISGRO ranges from 0.611s to 64.36s, and the other methods
are considerably slower.As shown in Figure 4B, the average time costs
of KarmaLoop on the CASP13+14 and CASP15 datasets are 0.047s and 0.049s,
respectively, displaying a considerable speed advantage over traditional
methods.Specifically, the minimum speed advantages of KarmaLoop over
DISGRO (17.329s and 20.526s) are 368× and 418× for CASP13+14 and CASP15,

\includegraphics[width=4.8in]{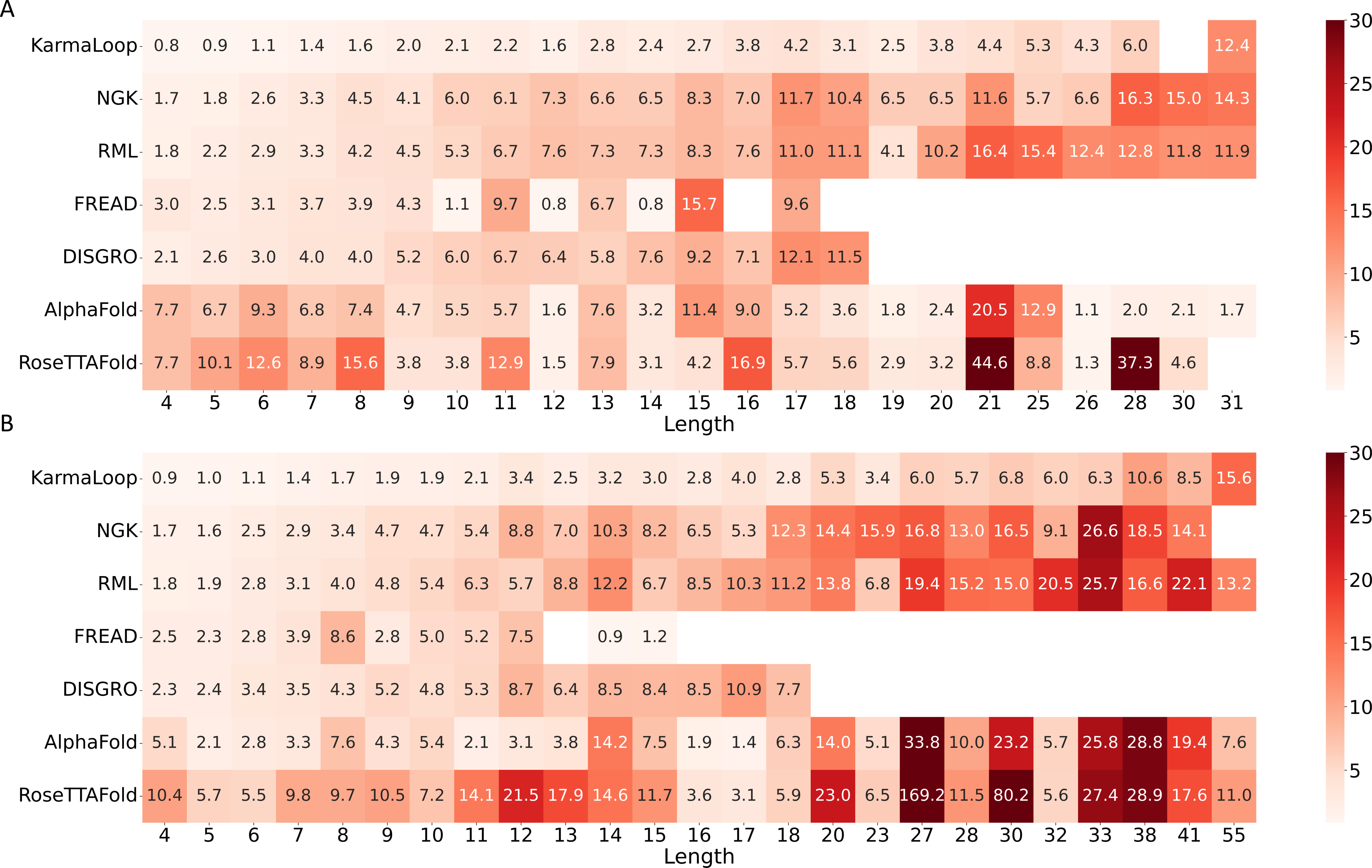}

{\footnotesize Figure 4. The time cost of each tested method. (A) Time
distribution of each method. (B) Average time cost. Further details on
the hardware setup can be found in the main text.\\}

respectively, and the maximum speed advantages of KarmaLoop over RML
(1433.562s and 945.091s) are 30,489× and 19,287×, respectively. However,
it should be noted that unlike other methods that only model loop
region, AlphaFold and RoseTTAFold require predicting the entire protein
conformation. Thus, the modeling speed for these methods cannot be
directly compared. While the reported speedup may vary slightly
depending on the hardware used, the two to four orders of magnitude in
speedup should still hold unless these baseline tools have been
completely re-engineered to boost their efficiency.

\textbf{\large 2.4  Prediction of antibody H3 loops}

The CDR H3 of antibody is widely acknowledged as the most flexible
region\textsuperscript{59} and contributes the most to the structural
diversity and binding site topography\textsuperscript{16}. Consequently,
accurate modeling of the H3 conformation is urgently needed. Sphinx,
proposed by Marks et al., is a hybrid method that models both general
loops and antibody H3 loops with the same architecture but using
different fragment databases\textsuperscript{22}. In contrast, KarmaLoop
consistently employs the same process for H3 modeling.

In this section, we present a comparative analysis of the H3 modeling
performance of KarmaLoop against four widely used methods (i.e.,
DeepAb\textsuperscript{42}, RosettaAntibody-G\textsuperscript{60},
RepertoireBuilder\textsuperscript{61}, and
AbodyBuilder\textsuperscript{62}) for predicting antibody structures on
the antibody H3 benchmark. To evaluate the quality of the modeling
results, we used the experimentally determined structures from SAbDab as
the input. Figure 5A depicts the distribution of the RMSDs between the
predicted samples from the baseline methods and the experimentally
determined structures.The results demonstrate that KarmaLoop (2.09 Å,
2.24 Å) outperforms DeepAb 
\includegraphics[width=4.8in,height=4.375in]{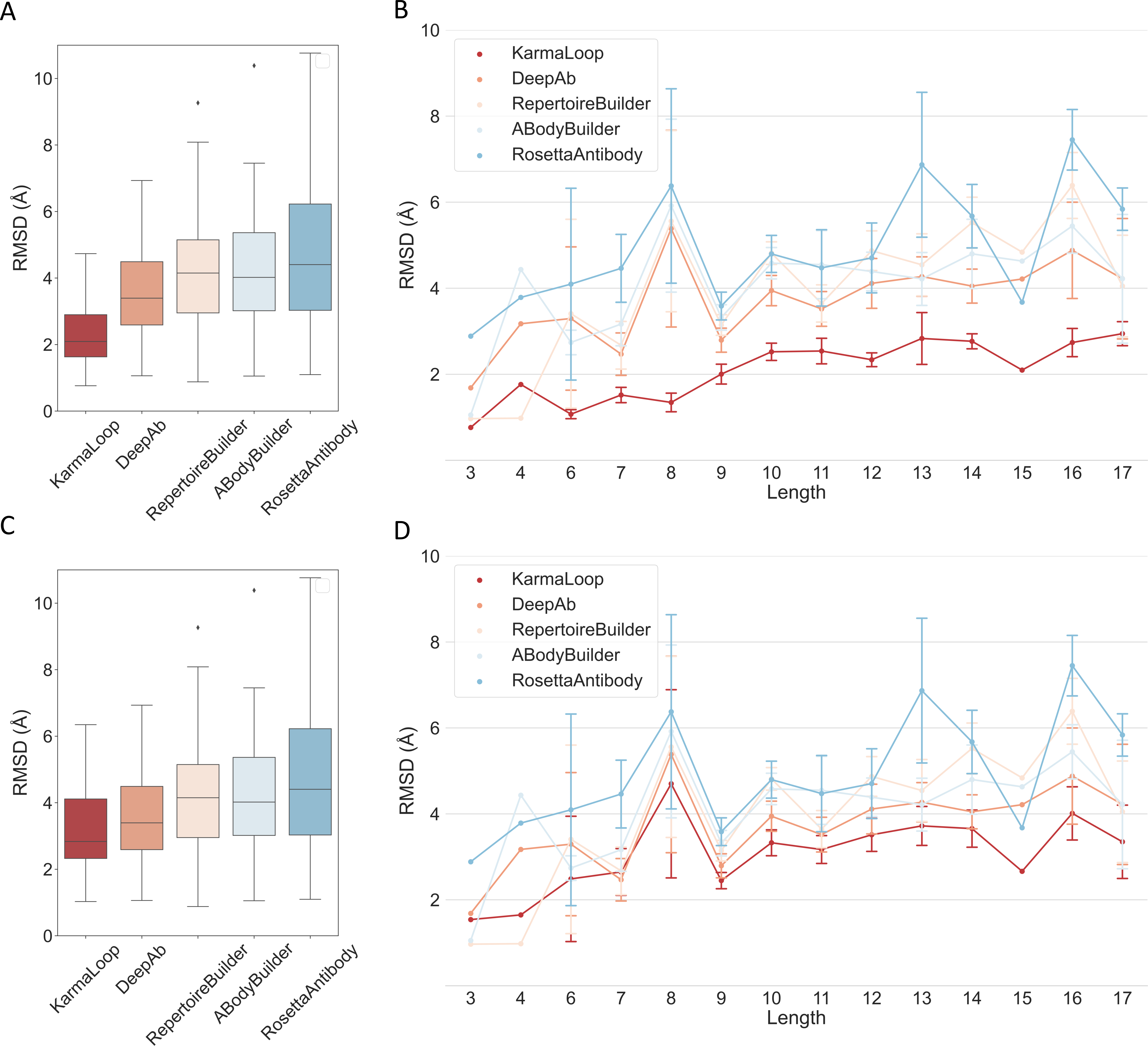}

{\footnotesize Figure 5. Comparison of the antibody CDR H3 loop prediction accuracy.
(A) Distribution of the RMSD values for each method. (B) Average RMSDs
of different lengths on the tested methods. The error bar shows the Std.
of the corresponding length target. (C) Distribution of the RMSD values
for each method (KarmaLoop makes refinement based on the antibody
structures modeled by DeepAb). (D) Average RMSDs of different length on
the tested methods (KarmaLoop makes refinement based on the antibody
structures modeled by DeepAb).\\}

(3.40 Å, 3.73 Å) in terms of the medium and
average RMSDs by 1.31 Å and 1.49 Å, respectively.  Figure 5B displays the
RMSDs of samples with varying loop lengths, and KarmaLoop exhibits the
lowest average RMSD for almost all lengths of H3 loop samples.

Next, we assessed the ability of refinement by using the
DeepAb-predicted antibody structures as the input for KarmaLoop. Figure
5C\&D illustrate the predictions based on the structures modeled by
DeepAb, which can be used to test the capability of KarmaLoop to improve
structures under a suboptimal structural situation. The results depicted
in Figure 5C\&D demonstrate that KarmaLoop is capable of refining the
loop structures with an average RMSD reduction from 3.73 Å to 3.24 Å and
a medium RMSD reduction from 3.40 Å to 2.81 Å, resulting in 13.14\% and
17.35\% improvement, respectively. It is worth noting that this
improvement is observed across most lengths of the H3 loops,
highlighting the consistent ability of KarmaLoop to improve the H3 loop
structures regardless of their length. Thus, a universal protocol can be
applied for antibody prediction, whereby DeepAb models the entire
antibody structures and KarmaLoop reconstructs the H3 structures.

\textbf{\large 2.5  Learning the principle of physical interactions}

To elucidate the interpretability of KarmaLoop, we present a pair of
case studies by portraying both covalent and non-covalent interactions.
As exhibited in Figure 6A, the attention values between the start nodes
and end nodes are displayed, which demonstrates the proficiency of
KarmaLoop in apprehending the interactions between the anchor residues
and the adjacent loop termini residues, a critical facet for the
formation of loop conformation\textsuperscript{22}. Figure 6B, on the
other hand, illustrates the capacity of KarmaLoop to perceive
non-covalent interactions, such as $\pi$-stacking, which are important in
the genesis of internal loop conformation. Overall, these findings
suggest that KarmaLoop is capable of capturing implicit interactions
without the need for prior knowledge.

\includegraphics[width=4.8in,height=2in]{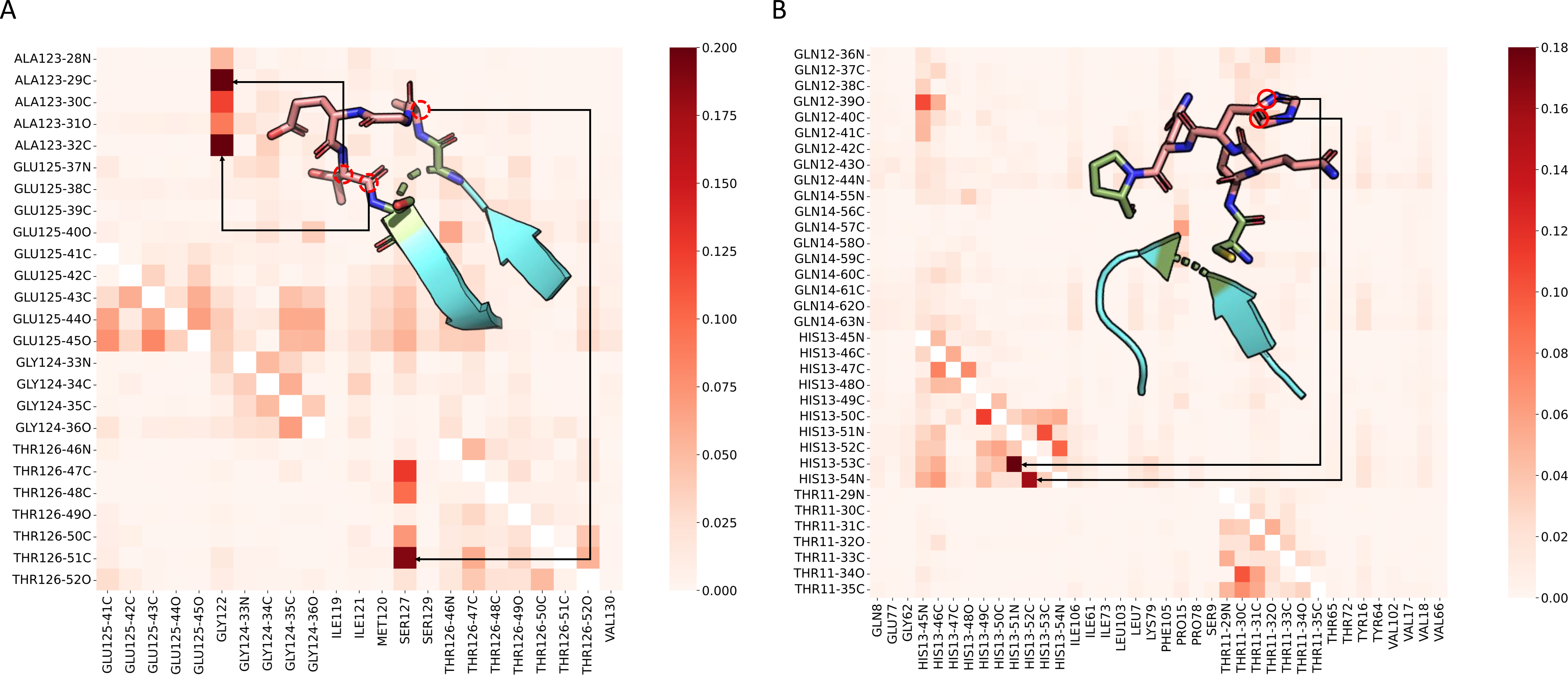}

{\footnotesize Figure 6. Attention weights in KarmaLoop. (A) Interactions
between anchor residues (before or after loop region) and adjacent loop
termini residues, the attention values increase with red gets deeper,
pink represents the loop residues, green represents the anchor residues,
blue is the non-loop and non-anchor region, and the red dashed circle
indicates the atoms of loop residue that have interactions with anchor
residues (PDB ID: T1071-D1 from CASP15). (B) Interactions of internal
bonds of loop residues, the red circle indicates the bond between two
atoms (PDB ID: T1066 from CASP14).\\}

\textbf{\Large 3 Conclusion}

In this work, we introduced a novel DL paradigm, KarmaLoop, to predict
and model protein loop conformations. KarmaLoop achieved sub-angstrom
accuracy and high efficiency based on the following innovative
components: (1) a hierarchical representation of residues and atoms to
precisely characterize protein structures; (2) the use of MDN for
learning pair-wise distances between loop and non-loop nodes to
facilitate loop conformation generation; (3) the deployment of
self-attention-based EGNN with fully-connected interaction graphs for
fast coordinate prediction; (4) an integrated clash correction procedure
to enforce the generation of chemically plausible conformations.

To verify the effectiveness of KarmaLoop, several evaluations and
comparisons have been conducted, including general loop modeling
accuracy, loop-length impact and compute time, as well as potential
applications in antibody H3 prediction. Of note, KarmaLoop achieved a
significant improvement in the average RMSD by more than two-fold and
the medium RMSD by nearly two-fold. Regarding the success rate, the
minimum enhancement of 28.09\% and 19.90\% in the CASP13+14 and CASP15
benchmark datasets were observed on one conformation generated by each
method. Even when compared to other methods that generate multiple
conformations, KarmaLoop with only one generated conformation still
achieved a minimum improvement of 13.27\% and 9.94\%. KarmaLoop
is also capable of accurately modeling loops of different lengths, and
obtained a majority length preponderance against the other methods.
Surprisingly, KarmaLoop achieved at least a 368× speed advantage than
other traditional methods. In term of antibody CDR H3 loop modeling,
which is considered the most challenging task in antibody structure
prediction, KarmaLoop also showed the best performance and maintained
its superiority across various lengths. Notably, based on the
DeepAb-modeled antibody structures, KarmaLoop has successfully promoted
its accuracy and achieved improvement upon DeepAb in each length of H3
loop.

In summary, KarmaLoop has demonstrated the potential to accurately model
loops, refine loop conformations based on inaccurate structure
environments, and significantly accelerate batch data processing.
Besides, a universal protocol was proposed for antibody structure
prediction by utilizing DeepAb to model the entire structure of the
antibody and KarmaLoop to refine the H3 conformation. Except for an
accurate loop modeling paradigm proposed, a large training dataset (0.27
million) and several benchmark datasets have been made available for
researchers to further evaluate the prediction accuracy of the general
loop and antibody H3. We anticipate that KarmaLoop will accelerate the
advancement of protein design, antibody-antigen recognition and drug
discovery, and will become an essential tool in the field of biology
moving forward.

Despite the considerable advancements achieved, KarmaLoop continues to
exhibit certain limitations. The model currently only provides a single
conformation, and there is a discernible decrease in accuracy when
confronted with extremely long loops. To address these constraints in
future development, two potential directions can be explored. First,
designing a probabilistic model capable of generating multiple
conformations might expand the competencies of model. Second, enriching
the training data with additional long-loop samples could augment the
model's efficacy in handling complex long-loop tasks. These enhancements
represent promising avenues for subsequent investigations.\\\\

\textbf{\Large 4  Method}

\textbf{\large 4.1  Architecture overview}

Inspired by Jin\textsuperscript{63}, we consider the loop modeling
problem as the docking task and selected any residues within 12 Å of
loop atoms as the loop pocket, which is a common and frequently used
protocol for ligand docking\textsuperscript{53}. As depicted in Figure
1B, the entire protein pocket is characterized as a 2D molecular graph
\(G_{p}\), with atoms as nodes and covalent bonds as edges. The non-loop
region of the protein is represented as a 3D residue graph \(G_{nl}\)
with residues as nodes and edges connecting the top 30 nearest neighbor
nodes, which captures long-range interactions better and has a lower
computational cost than a graph with atoms as nodes.

Instead of directly being used as the inputs for downstream tasks, the
molecular graphs and residue graphs are encoded by Graph
Transformer\textsuperscript{53-55} (GT) and Geometric Vector
Perceptrons\textsuperscript{56} (GVP) to learn intra-molecular
interactions and update node embeddings. Then, for non-loop residues,
the node embeddings for atom nodes in the same residue are summed and
concatenated to the corresponding residue node embeddings, followed by a
feature merging block (FMB) to generate hierarchical node embeddings.

\begin{equation}
H_{p,s}^{1} = GT\left( H_{p,s}^{0},E_{p,s}^{0} \right)
\end{equation}

\begin{equation}
H_{nl,s}^{0.5} = GVP\left( {S_{nl,s}^{0},H}_{nl,s}^{0},{H_{nl,v}^{0},E_{nl,s}^{0},E}_{nl,v}^{0} \right)
\end{equation}

\begin{equation}
H_{nl,s,i}^{1} = FMB\left( concat\left( H_{nl,s,i}^{0.5},\sum_{j = k}^{k'}H_{p,s,j}^{1} \right) \right)
\end{equation}

\begin{equation}
H_{l,s}^{1} = H_{p,s}^{1}\lbrack loop\_ mask\rbrack
\end{equation}

where \emph{H} and \emph{E} represent nodes and edges features,
respectively; \emph{S} represents the residue type; loop mask is a bool
array for selecting loop atoms from the whole protein atoms; \emph{p},
\emph{nl}, \emph{l}, \emph{s}, \emph{v}, \emph{i}, \emph{j}, \emph{k},
\emph{k'} on the lower right (i.e. subscripts) denote protein pocket
atoms, non-loop residues, loop residues, scalar, vector, the index of
residues, the index of atoms, the start index of the atom corresponding
to \emph{i}th residue, and the end index of the atom corresponding to
\emph{i}th residue, respectively.

Then, the node embeddings \(\left( H_{l,s}^{1},H_{nl,s}^{1} \right)\)
from the encoders and the node coordinates
\(\left( X_{l,s}^{1},X_{nl,s}^{1} \right)\) of both the loop atoms and
non-loop residues are combined to form an interaction graph
\(\left\lbrack G_{l,nl} = \left( H_{l,nl},\ E_{l,nl},X_{l,nl} \right) \right\rbrack\),
which captures inter-molecular interactions at both the residue-atom and
atom-atom levels.

In the process of pose generation, the loop conformation, represented as
the coordinates of the loop atom nodes, is initialized by sampling from
the normal distribution with a mean around the pocket center and a
standard deviation of 4 Å. The node embeddings and edge features are
initialized by graph normalization and an MLP layer (formula 5). Then
the node embeddings, edge embeddings and loop atoms positions are
updated through the application of the EGNN block consisting of 8 EGNN
layers with self-attention, taking into account both interactions
between loop, non-loop and loop, loop interactions (equation 6).
Inspired by AlphaFold, a recycling strategy is employed to enable the
EGNN block to learn how to consistently refine the poses. At the start
of each recycling, the updated embeddings and the raw embeddings will be
well combined through a gate block (equation 7).

\begin{equation}
H_{l,nl}^{0,0} = GraphNorm\left( H_{l,s},\ H_{nl,s} \right)
\end{equation}

\begin{equation}
H_{l,nl}^{r,i},E_{l,nl}^{r,i},X_{l,nl}^{r,i} = {EGNN\_ Layer}^{i}\left( H_{l,nl}^{r,i - 1},E_{l,nl}^{r,i - 1},X_{l,nl}^{r,i - 1}\  \right)
\end{equation}

\begin{equation}
F_{l,nl}^{r + 1,0} = Gate\_ Block\left( F_{l,nl}^{r,8},\ F_{l,nl}^{0,0} \right)
\end{equation}

where \emph{H} and \emph{E} represent embeddings of node and edge,
respectively; \emph{F} refers to an embedding in general; \emph{l},
\emph{nl}, \emph{r} and \emph{i} represent loop atoms, non-loop
residues, the recycling index and the EGNN layer index, respectively.

Upon obtaining node embeddings, the MDN block is used to predict
essential statistical parameters such as the mean, standard deviation,
and coefficient of variation. These parameters collectively provide a
comprehensive representation of the inter-nodal distance distribution,
as outlined in equation 8. Following this step, the node-pair embeddings
and inter-nodal distances are integrated with the predicted distribution
to ascertain the likelihood that an inter-nodal distance mirrors that
observed in the crystal conformation (formula 9). This resultant
probability estimation can then be utilized as a scoring metric to
evaluate the appropriateness of the current conformation in subsequent
analyses.

\begin{equation}
\mu_{l,nl},\sigma_{l,nl},\pi_{l,nl} = MDN\_ Block\left( H_{l,nl}^{1} \right)
\end{equation}

\begin{equation}
U_{(x)} = - \sum_{nl = 1}^{NL}{\sum_{l = 1}^{L}{\log P\left( \left( d_{l,nl} \middle| h_{l},h_{nl} \right) \right)}} = - score    
\end{equation}

where \emph{H} represents the node embeddings; \emph{l}, \emph{nl}
denote loop atoms and non-loop residues respectively;
\(\mu_{l,nl},\sigma_{l,nl},\pi_{l,nl}\) represent the means, standard
deviations and mixing coefficients, respectively.

\textbf{\large 4.2  Post-processing}

After KarmaLoop predicted the conformation, an optional Force Field (FF)
optimization was implemented to remove unfavorable steric clashes and
obtain high-quality loop conformations. OpenMM\textsuperscript{64} was
deployed for energy minimization by using the ff14SB force field to
optimize the predicted conformation. To prevent significant changes to
the protein, we applied a harmonic potential that restrains the backbone
atoms (N, CA, C, and CB) to their original locations. Finally, we ran
energy minimization on a GPU and saved the optimized protein structure
to a new PDB file.

\textbf{\large 4.3  Benchmark dataset}

The biennial Critical Assessment of protein Structure Prediction (CASP)
is a worldwide contest for
\href{https://en.wikipedia.org/wiki/Protein_structure_prediction}{protein
structure prediction}, and is also commonly used for loop modeling
assessment\textsuperscript{17,20,23,65}. The general loop benchmark
consists of two datasets: CASP13+14 and CASP15. CASP13+14 is the
combination of CASP13 and CASP14, which contains 549 loop structures
from 52 proteins (the unmatched structure and sequence sample was
removed). CASP15 containing 430 loop structures from 48 proteins was
collected from the latest CASP competition. The antibody H3 benchmark
derived from DeepAb is the combination of the RosettaAntibody and
therapeutic benchmark\textsuperscript{66}, and comprises 92 antibody CDR
H3 loop structures from 92 antibodies.

\textbf{\large 4.4  Training Dataset}

In this study, the training data was collected from PDB and Structural
Antibody Database (SAbDab)\textsuperscript{67}. Th protein data from PDB
was filtered by PISCES\textsuperscript{68} using the following criteria:
solved by X-ray crystallography, sequence identity $\leq$90\%, resolution
\textless3.0 Å, R-factor$\leq$0.25, and containing no DNA, RNA and UNK
molecules. The non-secondary structure of PDB data was defined by
DSSP\textsuperscript{69} and we defined loops as regions connecting two
secondary structures that consist of at least four residues. The
antibody dataset was selected to be the same as that of DeepAb. The
training and validation datasets were created using random stratified
sampling according to the loop length in a 9:1 ratio. The antibody data
was annotated using the Chothia numbering scheme\textsuperscript{70}.
Specifically, the training data contain the CDR L1, L2, L3 H1, H2 and H3
regions, and each type of CDRs was also split following the 9:1 ratio,
according to loop length. Finally, the total numbers of the data for
training and validation are 250,257 and 27,832, respectively.

\textbf{\large 4.5  Dataset preprocessing}

The protein data was downloaded from PDB, SAbDab and CASP. Water
molecules were subsequently removed from the proteins, and the loop
regions with non-standard amino acid residues or missing residues were
removed. The pocket region was defined as the non-loop residues within
12 Å of the loop atoms. Selecting the pocket instead of the entire
protein can simplify the input of network and guide KarmaLoop to focus
on the relevant structural features that are essential for understanding
the interaction between loop and non-loop regions. Then, we initialized
loop conformation by sampling from a normal distribution around the loop
region.

\textbf{\large 4.6  Evaluation methods}

In order to evaluate the performance of KarmaLoop, various widely-used
traditional and DL methods were selected for comparison. The general
protein loop modeling methods that were considered include (1)
knowledge-based method: FREAD; (2) \emph{ab initio} methods: DISGRO,
NGK, and Rosetta-missing-loop (RML); (3) DL-based methods: AlphaFold and
RoseTTAFold. According to the user guide, NGK requires a start
conformation of loop region and if there is a missing loop, and RML can
be used to fill it. In addition to these methods, several antibody
structure prediction methods were also examined including DeepAb,
RosettaAntibody-G\textsuperscript{60},
RepertoireBuilder\textsuperscript{61}, and
AbodyBuilder\textsuperscript{62} to assess the prediction ability of H3
loop region.

\textbf{\large 4.7  Metrics}

For loop modeling tasks, it is essential to consider both accuracy and
efficiency. Full-atom RMSD was utilized for evaluating the accuracy of
prediction. KarmaLoop and traditional methods (FREAD, DISGRO, NGK and
RML) focus solely on predicting the loop region of proteins, the RMSD
values were then calculated directly based on the predicted
conformations. Conversely, AlphaFold and RoseTTAFold model the entire
protein structures, so that the predicted structures of protein are
aligned before calculating RMSD values. Additionally, DeepAb,
RosettaAntibody-G, RepertoireBuilder, and AbodyBuilder and
RosettaAntibody predict the complete structure of antibodies. Following
the same guidelines with DeepAb, the heavy chains of these antibodies
are aligned prior to the RMSD calculation. Moreover, the success rate of
the method was introduced as a complementary metric to RMSD, which
measures the proportion of cases where the RMSD is less than or equal to
2 Å compared to experimentally determined conformations. The efficiency
was simply measured by computing time.

\textbf{\large 4.8  Computing resource}

KarmaLoop was trained on 8 NVIDIA A100-SXM4-80GB and 64 cores Intel(R)
Xeon(R) Platinum 8358P CPU @ 2.60GHz. For evaluation, KarmaLoop was
evaluated on a Tesla V100S GPU. Traditional methods (FREAD, DISGRO, NGK
and RML) were accomplished in parallel with 48 cores Intel(R) Xeon(R)
Gold 6240R CPU @ 2.40GHz. AlphaFold and RoseTTAFold were executed in
parallel on 20 cores of the Intel(R) Xeon(R) Gold 6240R CPU @ 2.40GHz,
and a Tesla V100S GPU.\\

\textbf{\Large Data and Code Availability}

The source code will be
available at \url{https://github.com/karma211225/KarmaLoop} once the paper is accepted. \\

\textbf{\Large Acknowledgments}

This study was supported by the National Key Research and Development
Program of China (2022YFF1203000), the National Natural Science
Foundation of China (22220102001, 82204279, 22007082, 62006219), the
Fundamental Research Funds for the Central Universities
(226-2022-00220), Natural Science Foundation of Zhejiang Province
(LQ21B030013) and Hong Kong Innovation and Technology Fund (Project No.
ITS/241/21).\\

\textbf{\Large References}

1 Tang, K., Zhang, J. \& Liang, J. Fast protein loop sampling and
structure prediction using distance-guided sequential chain-growth Monte
Carlo method. \emph{PLoS Comput Biol} \textbf{10}, e1003539 (2014).

2 Rufino, S. D., Donate, L. E., Canard, L. H. \& Blundell, T. L.
Predicting the conformational class of short and medium size loops
connecting regular secondary structures: application to comparative
modelling. \emph{J Mol Biol} \textbf{267}, 352-367 (1997).

3 Lins, L., Thomas, A. \& Brasseur, R. Analysis of accessible surface of
residues in proteins. \emph{Protein Science} \textbf{12}, 1406-1417
(2003).

4 Stevens, A. O. \& He, Y. Benchmarking the Accuracy of AlphaFold 2 in
Loop Structure Prediction. \emph{Biomolecules} \textbf{12} (2022).

5 Lee, J., Lee, D., Park, H., Coutsias, E. A. \& Seok, C. Protein loop
modeling by using fragment assembly and analytical loop closure.
\emph{Proteins} \textbf{78}, 3428-3436 (2010).

6 Li, Z. \emph{et al.} Flexibility Regulation of Loops Surrounding the
Tunnel Entrance in Cytochrome P450 Enhanced Substrate Access
Substantially. \emph{ACS Catalysis} \textbf{12}, 12800-12808 (2022).

7 Crean, R. M., Biler, M., van der Kamp, M. W., Hengge, A. C. \&
Kamerlin, S. C. L. Loop Dynamics and Enzyme Catalysis in Protein
Tyrosine Phosphatases. \emph{Journal of the American Chemical Society}
\textbf{143}, 3830-3845 (2021).

8 Borges, P. T. \emph{et al.} Methionine-Rich Loop of Multicopper
Oxidase McoA Follows Open-to-Close Transitions with a Role in Enzyme
Catalysis. \emph{ACS Catalysis} \textbf{10}, 7162-7176 (2020).

9 Jones, S. \& Thornton, J. M. Prediction of protein-protein interaction
sites using patch analysis11Edited by G. von Heijne. \emph{J Mol Biol}
\textbf{272}, 133-143 (1997).

10 Streaker, E. D. \& Beckett, D. Ligand-linked Structural Changes in
the Escherichia coli Biotin Repressor: The Significance of Surface Loops
for Binding and Allostery. \emph{J Mol Biol} \textbf{292}, 619-632
(1999).

11 Myllykoski, M., Raasakka, A., Han, H. \& Kursula, P. Myelin
2',3'-cyclic nucleotide 3'-phosphodiesterase: active-site ligand binding
and molecular conformation. \emph{PLoS One} \textbf{7}, e32336 (2012).

12 Tamamis, P. \& Floudas, Christodoulos~A. Molecular Recognition of
CXCR4 by a Dual Tropic HIV-1 gp120 V3 Loop. \emph{Biophysical Journal}
\textbf{105}, 1502-1514 (2013).

13 Papaleo, E. \emph{et al.} The Role of Protein Loops and Linkers in
Conformational Dynamics and Allostery. \emph{Chem Rev} \textbf{116},
6391-6423 (2016).

14 Zhao, S., Zhu, K., Li, J. \& Friesner, R. A. Progress in super long
loop prediction. \emph{Proteins} \textbf{79}, 2920-2935 (2011).

15 Berman, H. M. \emph{et al.} The Protein Data Bank. \emph{Nucleic
Acids Res} \textbf{28}, 235-242 (2000).

16 Marks, C. \& Deane, C. M. Antibody H3 Structure Prediction.
\emph{Comput Struct Biotechnol J} \textbf{15}, 222-231 (2017).

17 Choi, Y. \& Deane, C. M. FREAD revisited: Accurate loop structure
prediction using a database search algorithm. \emph{Proteins}
\textbf{78}, 1431-1440 (2010).

18 Messih, M. A., Lepore, R. \& Tramontano, A. LoopIng: a template-based
tool for predicting the structure of protein loops.
\emph{Bioinformatics} \textbf{31}, 3767-3772 (2015).

19 Hildebrand, P. W. \emph{et al.} SuperLooper-\/-a prediction server
for the modeling of loops in globular and membrane proteins.
\emph{Nucleic Acids Res} \textbf{37}, W571-W574 (2009).

20 Karami, Y., Guyon, F., De Vries, S. \& Tufféry, P. DaReUS-Loop:
accurate loop modeling using fragments from remote or unrelated
proteins. \emph{Sci Rep} \textbf{8}, 13673 (2018).

21 Barozet, A., Chacón, P. \& Cortés, J. Current approaches to flexible
loop modeling. \emph{Curr Res Struct Biol} \textbf{3}, 187-191 (2021).

22 Marks, C. \emph{et al.} Sphinx: merging knowledge-based and ab initio
approaches to improve protein loop prediction. \emph{Bioinformatics}
\textbf{33}, 1346-1353 (2017).

23 Liang, S., Zhang, C. \& Zhou, Y. LEAP: highly accurate prediction of
protein loop conformations by integrating coarse-grained sampling and
optimized energy scores with all-atom refinement of backbone and side
chains. \emph{J Comput Chem} \textbf{35}, 335-341 (2014).

24 Canutescu, A. A. \& Dunbrack, R. L. Cyclic coordinate descent: A
robotics algorithm for protein loop closure. \emph{Protein Sci}
\textbf{12}, 963-972 (2003).

25 Wang, C., Bradley, P. \& Baker, D. Protein-protein docking with
backbone flexibility. \emph{J Mol Biol} \textbf{373}, 503-519 (2007).

26 Mandell, D. J., Coutsias, E. A. \& Kortemme, T. Sub-angstrom accuracy
in protein loop reconstruction by robotics-inspired conformational
sampling. \emph{Nature methods} \textbf{6}, 551-552 (2009).

27 Stein, A. \& Kortemme, T. Improvements to robotics-inspired
conformational sampling in rosetta. \emph{PLoS One} \textbf{8}, e63090
(2013).

28 Deane, C. M. \& Blundell, T. L. CODA: a combined algorithm for
predicting the structurally variable regions of protein models.
\emph{Protein Sci} \textbf{10}, 599-612 (2001).

29 Deane, C. M. \& Blundell, T. L. A novel exhaustive search algorithm
for predicting the conformation of polypeptide segments in proteins.
\textbf{40}, 135-144 (2000).

30 Deng, Y. \emph{Deep learning on mobile devices: a review}. Vol. 10993
DCS (SPIE, 2019).

31 Jumper, J. \emph{et al.} Highly accurate protein structure prediction
with AlphaFold. \emph{Nature} \textbf{596}, 583-589 (2021).

32 Baek, M. \emph{et al.} Accurate prediction of protein structures and
interactions using a three-track neural network. \textbf{373}, 871-876
(2021).

33 Lee, C., Su, B.-H. \& Tseng, Y. J. Comparative studies of AlphaFold,
RoseTTAFold and Modeller: a case study involving the use of
G-protein-coupled receptors. \emph{Brief Bioinform} \textbf{23} (2022).

34 Liang, T. \emph{et al.} Differential performance of RoseTTAFold in
antibody modeling. \emph{Brief Bioinform} (2022).

35 LeCun, Y. \emph{et al.} Backpropagation Applied to Handwritten Zip
Code Recognition. \emph{Neural Computation} \textbf{1}, 541-551 (1989).

36 Krizhevsky, A., Sutskever, I. \& Hinton, G. E. ImageNet
classification with deep convolutional neural networks. \textbf{60},
84--90 (2017).

37 Zhang, N., Donahue, J., Girshick, R. \& Darrell, T. 834-849 (Springer
International Publishing).

38 Ouyang, W. \emph{et al.} DeepID-Net: multi-stage and deformable deep
convolutional neural networks for object detection. arXiv:1409.3505
(2014).
\textless{}\url{https://ui.adsabs.harvard.edu/abs/2014arXiv1409.3505O}\textgreater.

39 He, K., Zhang, X., Ren, S. \& Sun, J. in \emph{IEEE Conference on
Computer Vision and Pattern Recognition.}

40 Ruffolo, J. A., Guerra, C., Mahajan, S. P., Sulam, J. \& Gray, J. J.
Geometric potentials from deep learning improve prediction of CDR H3
loop structures. \emph{Bioinformatics} \textbf{36}, i268-i275 (2020).

41 Gers, F. A., Schmidhuber, J. \& Cummins, F. Learning to Forget:
Continual Prediction with LSTM. \emph{Neural Computation} \textbf{12},
2451-2471 (2000).

42 Ruffolo, J. A., Sulam, J. \& Gray, J. J. J. P. Antibody structure
prediction using interpretable deep learning. \textbf{3}, 100406 (2022).

43 Abanades, B., Georges, G., Bujotzek, A. \& Deane, C. M. ABlooper:
fast accurate antibody CDR loop structure prediction with accuracy
estimation. \emph{Bioinformatics} \textbf{38}, 1877-1880 (2022).

44 Garcia Satorras, V., Hoogeboom, E. \& Welling, M. J. a. e.-p. E(n)
Equivariant Graph Neural Networks. arXiv:2102.09844 (2021).
\textless{}\url{https://ui.adsabs.harvard.edu/abs/2021arXiv210209844G}\textgreater.

45 Liang, S., Zhang, C., Sarmiento, J. \& Standley, D. M. Protein Loop
Modeling with Optimized Backbone Potential Functions. \emph{J Chem
Theory Comput} \textbf{8}, 1820-1827 (2012).

46 Bonet, J., Segura, J., Planas-Iglesias, J., Oliva, B. \&
Fernandez-Fuentes, N. Frag'r'Us: knowledge-based sampling of protein
backbone conformations for de novo structure-based protein design.
\emph{Bioinformatics} \textbf{30}, 1935-1936 (2014).

47 López-Blanco, J. R., Canosa-Valls, A. J., Li, Y. \& Chacón, P. RCD+:
Fast loop modeling server. \emph{Nucleic Acids Res} \textbf{44},
W395-W400 (2016).

48 Chys, P. \& Chacón, P. Random Coordinate Descent with Spinor-matrices
and Geometric Filters for Efficient Loop Closure. \emph{J Chem Theory
Comput} \textbf{9}, 1821-1829 (2013).

49 Coutsias, E. A., Seok, C., Jacobson, M. P. \& Dill, K. A. A kinematic
view of loop closure. \emph{J Comput Chem} \textbf{25}, 510-528 (2004).

50 Gray, J. J. \emph{et al.} Protein--Protein Docking with Simultaneous
Optimization of Rigid-body Displacement and Side-chain Conformations.
\emph{J Mol Biol} \textbf{331}, 281-299 (2003).

51 Meiler, J. \& Baker, D. ROSETTALIGAND: Protein--small molecule
docking with full side-chain flexibility. \textbf{65}, 538-548 (2006).

52 Leach, A. R. Ligand docking to proteins with discrete side-chain
flexibility. \emph{J Mol Biol} \textbf{235}, 345-356 (1994).

53 Shen, C. \emph{et al.} Boosting Protein--Ligand Binding Pose
Prediction and Virtual Screening Based on Residue--Atom Distance
Likelihood Potential and Graph Transformer. \emph{Journal of Medicinal
Chemistry} \textbf{65}, 10691-10706 (2022).

54 Morehead, A., Chen, C. \& Cheng, J. J. a. e.-p. Geometric
Transformers for Protein Interface Contact Prediction. arXiv:2110.02423
(2021).
\textless{}\url{https://ui.adsabs.harvard.edu/abs/2021arXiv211002423M}\textgreater.

55 Dwivedi, V. P. \& Bresson, X. J. a. e.-p. A Generalization of
Transformer Networks to Graphs. arXiv:2012.09699 (2020).
\textless{}\url{https://ui.adsabs.harvard.edu/abs/2020arXiv201209699D}\textgreater.

56 Jing, B., Eismann, S., Suriana, P., Townshend, R. J. L. \& Dror, R.
J. a. e.-p. LEARNING FROM PROTEIN STRUCTURE WITH GEOMETRIC VECTOR
PERCEPTRONS. arXiv:2009.01411 (2020).
\textless{}\url{https://ui.adsabs.harvard.edu/abs/2020arXiv200901411J}\textgreater.

57 Kryshtafovych, A., Schwede, T., Topf, M., Fidelis, K. \& Moult, J.
Critical assessment of methods of protein structure prediction
(CASP)-Round XIII. \emph{Proteins} \textbf{87}, 1011-1020 (2019).

58 Kryshtafovych, A., Schwede, T., Topf, M., Fidelis, K. \& Moult, J.
Critical assessment of methods of protein structure prediction
(CASP)---Round XIV. \textbf{89}, 1607-1617 (2021).

59 Luo, S. \emph{et al.} Antigen-Specific Antibody Design and
Optimization with Diffusion-Based Generative Models.
2022.2007.2010.499510 (2022).

60 Jeliazkov, J. R., Frick, R., Zhou, J. \& Gray, J. J. Robustification
of RosettaAntibody and Rosetta SnugDock. \emph{PLoS One} \textbf{16},
e0234282 (2021).

61 Schritt, D. \emph{et al.} Repertoire Builder: high-throughput
structural modeling of B and T cell receptors. \emph{Molecular Systems
Design \& Engineering} \textbf{4}, 761-768 (2019).

62 Dunbar, J. \emph{et al.} SAbPred: a structure-based antibody
prediction server. \emph{Nucleic Acids Res} \textbf{44}, W474-W478
(2016).

63 Jin, W., Barzilay, R. \& Jaakkola, T. J. a. e.-p. Antibody-Antigen
Docking and Design via Hierarchical Equivariant Refinement.
arXiv:2207.06616 (2022).
\textless{}\url{https://ui.adsabs.harvard.edu/abs/2022arXiv220706616J}\textgreater.

64 Eastman, P. \emph{et al.} OpenMM 4: A Reusable, Extensible, Hardware
Independent Library for High Performance Molecular Simulation. \emph{J
Chem Theory Comput} \textbf{9}, 461-469 (2013).

65 Lee, G. R., Heo, L. \& Seok, C. Effective protein model structure
refinement by loop modeling and overall relaxation. \emph{Proteins}
\textbf{84 Suppl 1}, 293-301 (2016).

66 Raybould, M. I. J. \emph{et al.} Five computational developability
guidelines for therapeutic antibody profiling. \textbf{116}, 4025-4030
(2019).

67 Dunbar, J. \emph{et al.} SAbDab: the structural antibody database.
\emph{Nucleic Acids Res} \textbf{42}, D1140-D1146 (2013).

68 Wang, G. \& Dunbrack, R. L. PISCES: recent improvements to a PDB
sequence culling server. \emph{Nucleic Acids Res} \textbf{33}, W94-W98
(2005).

69 Touw, W. G. \emph{et al.} A series of PDB-related databanks for
everyday needs. \emph{Nucleic Acids Res} \textbf{43}, D364-D368 (2015).

70 Chothia, C. \emph{et al.} Conformations of immunoglobulin
hypervariable regions. \emph{Nature} \textbf{342}, 877-883 (1989).

\end{document}